# Inspection and Test Process Integration based on Explicit Test Prioritization Strategies

Frank Elberzhager[1], Alla Rosbach[1], Jürgen Münch[2], Robert Eschbach[1]

[1] Fraunhofer IESE, Fraunhofer Platz 1,
67663 Kaiserslautern, Germany
{frank.elberzhager, alla.rosbach, robert.eschbach}@iese.fraunhofer.de
[2] University of Helsinki, P.O. Box 68,
00014 Helsinki, Finland
juergen.muench@cs.helsinki.fi

**Abstract.** Today's software quality assurance techniques are often applied in isolation. Consequently, synergies resulting from systematically integrating different quality assurance activities are often not exploited. Such combinations promise benefits, such as a reduction in quality assurance effort or higher defect detection rates. The integration of inspection and testing, for instance, can be used to guide testing activities. For example, testing activities can be focused on defect-prone parts based upon inspection results. Existing approaches for predicting defect-prone parts do not make systematic use of the results from inspections. This article gives an overview of an integrated inspection and testing approach, and presents a preliminary case study aiming at verifying a study design for evaluating the approach. First results from this preliminary case study indicate that synergies resulting from the integration of inspection and testing might exist, and show a trend that testing activities could be guided based on inspection results.

## 1 Introduction

Quality assurance activities, such as inspection (i.e., static quality assurance) and testing (i.e., dynamic quality assurance) activities, are an essential part of today's software development in order to ensure software products of high quality. However, the costs for performing quality assurance activities can consume more than 50 percent of the overall development effort, especially for testing [7]. Moreover, it is often unclear how to systematically guide and focus testing activities.

Existing approaches to focusing testing activities are widely based on metrics such as size or complexity, gathered from the development of current or historical software products. Regarding inspection and testing activities, systematic integration is often missing. Inspection and testing activities are usually applied in sequence, i.e., in isolation, and do not exploit synergy effects such as reduced effort or the use of inspection results to guide testing activities.

This article presents an integrated inspection and testing approach that is able to guide testing activities based on inspection results. Parts of a system that are expected to be most defect-prone or defect types that are especially relevant can be prioritized based on defect data gathered during an inspection. In order to be able to conduct a focused testing activity, knowledge about relationships between inspections and testing is required. Such relationships are usually context-specific. Therefore, it is necessary to prove whether reliable evidence about such relationships exists in a given context. If no such evidence exists, assumptions need to be made regarding the relationships between inspection and testing activities. For example, one assumption might be that parts of a system where a significant number of defects are found by an inspection contain more defects to be found by testing (i.e., a Pareto distribution of defects is expected). Such assumptions have to be evaluated in a given context in order to provide appropriate guidance for testing.

A study design was determined and verified during a preliminary case study in which the integrated inspection and testing approach was applied. The results showed, for instance, that inspection and testing activities should focus on defect types that are most suitable for detection (e.g., maintainability problems during inspections, and usability problems during testing), and that an effort reduction for testing of up to 23% was achievable in the given context. However, one important prerequisite for the applicability of the approach is the testability of the software under test.

The remainder of this article is structured as follows: Section 2 presents a short overview of the integrated inspection and testing approach. The study design and the preliminary case study are described in Section 3. Finally, Section 4 concludes the article and gives an outlook on future work. An extended version of this article includes related work and a more detailed description of the approach [5].

## 2 Approach

The main idea of the integrated inspection and testing approach [2], [3] is to use inspection defect data to guide testing activities. In doing so, parts of a system under test that are expected to be most defect-prone or defect types that are expected to show up during testing can be prioritized based on an inspection defect profile (consisting of, for example, quantitative defect data and defect type information from an inspection). The approach is able to prioritize parts of a system or defect types (1-stage approach), or both (2-stage approach), and can thus define a test strategy.

In order to be able to focus testing activities, it is necessary to describe relationships between defects found in the inspection and the remaining defect distribution in the system under test, which also counts for defect types. Consequently, assumptions are explicitly defined. One example of an assumption is that for parts of a system where many inspection defects are found, more defects are expected to be found with testing activities (i.e., a Pareto distribution of defects is expected). Assumptions should be at least grounded on explicitly described hypotheses to make them reliable. Nevertheless, each assumption has to be validated in a given environment in order to be able to decide whether the assumption can be accepted or not (and thus, checking whether valuable guidance for testing activities is provided).

In addition, context factors have to be considered, such as the number of available inspectors or the experience of the inspectors. For example, consider the number of available inspectors and time as two context factors. If only one inspector is available for inspecting certain parts of a system within a limited amount of time, fewer parts can be inspected. Consequently, more effort should be expended on testing activities.

Since an assumption is often too coarse-grained to be applied directly, concrete selection rules have to be derived in order to be operational. For example, the assumption regarding the Pareto distribution of defects can be refined in terms of application level and thresholds, leading to the following exemplary selection rule: "Focus a unit testing activity on code modules where the inspection found more than 10 major defects per 1,000 lines of code".

In addition to the inspection defect profile, metrics and historical data can be combined with inspection defect data in order to improve the prediction of defect-proneness and relevant defect types, and thus, to obtain improved guidance for testing activities. Fig. 1 presents the concrete process steps for guiding testing activities based on inspection results.

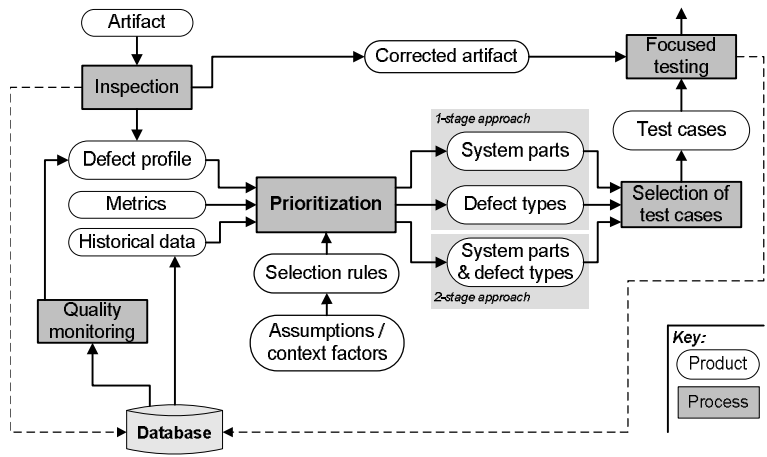

**Fig. 1.** Integrated inspection and testing approach

## 3  Case Study

### 3.1  Goals

The primary goals of the preliminary study were to check the design of the study and to evaluate whether an integrated inspection and testing approach is able to guide testing activities based on inspection results. Inspection defect data should be used to predict those parts of a system under test that remain especially defect-prone and should therefore be addressed by additional testing activities. In addition, defect types should be prioritized for testing based on inspection data.

While the initial case study showed first insights regarding the relationship between inspection and testing activities on the code level, more data should be gained in the preliminary case study presented here. Therefore, different assumptions in a given context describing the relationship between defects found during inspection and testing had to be evaluated regarding their suitability for guiding (i.e., focusing) testing activities.

The following research questions are derived from the primary goals:
- RQ1: Is an evaluation of the integrated approach possible with the given study design, and which assumptions between inspection and testing activities are most suitable in the given context for guiding testing activities?
    - RQ1.1: How appropriate is it to focus testing activities on certain parts of a product based on inspection defect results?
    - RQ1.2: How appropriate is it to focus testing activities on specific defect types based on inspection defect results?

### 3.2 Main Results from Another Evaluation

The preliminary case study presented in this article is similar to an earlier case study [2], [3]. It could be shown that assumptions regarding a Pareto distribution of defects led to suitable predictions of defect-prone parts, while combining inspection results and product metrics led to inconsistent results for the prediction of defect-prone code classes. The main differences of the preliminary study presented here compared to the previous one are that new assumptions are defined, that the integrated approach is evaluated in a different context (e.g., another product that was inspected and tested, new subjects), and that system testing is considered. Furthermore, the preliminary study presented here has a special focus on evaluating the design of the study before applying the approach in an industrial environment.

### 3.3 Context

A Java tool called DETECT (dependability focused inspection tool) was used for evaluating the integrated inspection and testing approach. The tool supports people who perform an inspection. Currently, it mainly supports individual defect detection with the help of different kinds of reading support and allows defining new checklists for use during defect detection. The different kinds of reading support include different tree structures and two kinds of checklists. The tool provides a three-part view for the inspector: a tracking mode that documents each step; the artifact to be checked; and the corresponding reading support (e.g., a checklist).

The tool was mainly developed by one developer. Currently, it consists of about 57k lines of code (without blank lines and comments), about 380 classes, and about

2,300 methods. The developer identified the critical code parts that should be inspected and discussed the selection of the code classes with the inspection team. In order to be able to finish the inspection within existing time constraints, it was decided to inspect only one kind of reading support, namely GITs (goal-indicator trees [6]). Overall, six inspectors checked 12 code classes, comprising about 7,300 lines of code. Each inspector checked four code classes, consisting of about 2,500 lines of code.

Table 1 shows the experience, respectively the knowledge, of the six inspectors regarding the inspection, the reading support to be checked, and the code structures (i.e., programming knowledge). Three values (low, middle, high) are used for the classification. Finally, the assigned checklists are shown.

The testing activities were performed by the developer of the tool and one additional tester. Neither one was involved in the inspection.

**Table 1.** Experience of inspectors and assigned checklists

| No. | Inspection knowledge | GIT knowledge | Programming knowledge | Defect detection focus |
|---|---|---|---|---|
| 1 | + | ++ | ++ | requirements |
| 2 | ++ | ++ | ++ | requirements |
| 3 | + | o | ++ | implementation |
| 4 | ++ | + | ++ | implementation, reliability |
| 5 | ++ | o | o | code documentation |
| 6 | ++ | ++ | + | code documentation |

### 3.4 Design

The preliminary study described in this article followed a similar design compared to the first evaluation of the integrated inspection and testing approach [2], [3].

First, a code inspection was conducted by six computer scientists (step 1) using checklists. Overall, four different checklists were used, addressing requirements fulfillment, implementation, reliability, and code documentation. Each checklist consisted of three to eight questions and was assigned to those inspectors who could answer the questions effectively. Finally, the checklists were mapped to the relevant code classes by the developer of the tool so that each inspector checked four code classes. One experienced quality assurance engineer aggregated the findings from all inspectors. The developer analyzed each problem and decided whether a real defect was found that had to be corrected or whether problems that were documented by an inspector were only due to a misunderstanding and could be removed without any correction.

The next step was the quality monitoring of the resulting inspection defect profile (step 2). Reading rate, overall number of found defects, and defect distribution were considered.

Step 3 comprised the prioritization, i.e., a prediction of defect-prone parts and defect types had to be conducted. For this, four context-specific assumptions were determined that were to be evaluated. Two assumptions of the initial study [2], [3] were reused, and two new assumptions were defined based on experiences made during the first study.

Finally, selecting test cases and conducting focused testing activities would be the last steps (step 4 and 5). However, in order to be able to evaluate the stated assumptions, prioritized as well as not prioritized parts were tested by two testers. This enabled a detailed analysis of the assumptions regarding their appropriateness. First, a unit test of code classes was started. Test cases were derived using equivalence partitioning. Code classes that had been inspected and some additional ones identified as being most critical or important were selected for testing. However, it turned out that efficient unit testing was not possible due to bad testability of the code classes. The code structure did not suit the unit test approach (e.g., due to anonymous inner classes, anonymous threads, private fields and methods). To neutralize the problems of the code structure, mocking frameworks (i.e., a simulation of the behavior of code classes) were used. However, this framework turned out to be very complex for inexperienced testers.

Beside unit testing, a manual system test was conducted in order to analyze whether prioritization is possible between different levels (i.e., using defect information from the code level to guide tests for the system level). System tests were derived through typical walkthrough scenarios that followed the main functionality the tool offers. Afterwards, the results from this testing activity were used as a baseline and compared to the prioritization when the defined assumptions were evaluated.

### 3.5 Execution

**Step 1: Performing the inspection**

Before the inspection was performed, a team meeting was held where the checklists were explained and an overview of the code to be inspected was presented. Afterwards, each inspector checked the assigned code classes with the assigned checklist and documented all findings and the place of occurrence in a problem list. In addition, defect type and defect severity were recorded. Each code class was checked by at least two inspectors. Overall, 1450 minutes were spent on individual defect detection (ranging from 90 to 280 minutes consumed per inspector).

**Table 2.** Defect content and defect density of each inspected code class

| Code class | I | II | III | IV | V | VI | VII | VIII | IX | X | XI | XII |
|---|---|---|---|---|---|---|---|---|---|---|---|---|
| Defect content | 4 | 18 | 19 | 2 | 34 | 18 | 13 | 24 | 31 | 11 | 10 | 5 |
| Defect density | .009 | .021 | .020 | .008 | .061 | .057 | .038 | .031 | .045 | .026 | .031 | .016 |

One experienced quality assurance engineer compiled the defect detection profile and the developer of the tool checked for each defect whether it has to be corrected or not. Of 236 problems found in total, 189 defects to be corrected remained. Table 2 shows the defect content (absolute number of defects) and defect density (absolute number of defects divided by lines of code) of the twelve inspected code classes. Table 3 shows a sorted list of the ODC-classified defects [8]. 54 defects (e.g., unclear or missing comments) could not be classified according to any of the existing defect types.

Table 3. ODC-classified defects from inspection

| ODC defect types | Sub-total | ODC defect types | Sub-total | Total |
|---|---|---|---|---|
| algorithm / method | 53 | relationship | 1 | |
| checking | 36 | timing / serialization | 0 | |
| function / class / object | 32 | interface / o-o messages | 0 | |
| assignment / initialization | 13 | other | 54 | |
| Sub-total | 134 | | 55 | **189** |

**Step 2: Monitoring the inspection results**

Because this was the first systematic quality assurance run of the DETECT tool, no historical data was available that could be used for monitoring the inspection results. Instead, data from the first study of the integrated inspection and testing approach [2], [3] was used since the environment was similar. In addition, data from the literature was considered. The reading rate was about 630 lines of code per hour, which is similar to the first study (there, it was 550, respectively 685, lines of code per hour in two quality assurance runs). The number is rather high compared to reading rates recommended in the literature, but consistent with experiences from industry [1]. Some reasons for the high number are that all lines of code were counted (including blank lines and comments) and that the individual checklists pinpointed the inspectors to certain parts, whereas other parts were read faster. Finally, the overall number of found defects seemed reasonable compared to the first study and the distribution of minor, major, and crash defects was also similar to the first study.

**Step 3: Prioritizing the testing activities**

In order to guide testing activities, a prediction of defect-prone parts and defects of those defect types that are expected to appear during testing was made, i.e., those parts and defect types were prioritized. Four assumptions were stated, including instructions for the prioritization. More details and explanations can be found in [2], [3], [4].

*Assumption 1:* If no defined selection criterion is used to determine parts of a system that should be inspected, it is expected that a significant number of defects still remain in those parts that are not inspected (i.e., an equal distribution of defects is assumed). Consequently, testing should be focused especially on those uninspected parts of a system.

*Assumption 2:* Parts of a system where a large number of inspection defects are found indicate more defects to be found with testing (i.e., a Pareto distribution of defects is assumed). Consequently, testing should be focused especially on those inspected parts of a system that were particularly defect-prone.

*Assumption 3:* Inspection and testing activities find defects of various defect types with different effectiveness. For inspections, this includes, e.g., maintainability problems. For testing, this includes, e.g., performance problems. Consequently, inspection and testing activities should be focused on those defect types that are most convenient to find.

*Assumption 4:* Defects of the defect types that are found most often by the inspection (i.e., a Pareto distribution of defects of certain defect types is assumed) indicate more defects of the defect types to be found with testing. Thus, testing should be focused on those defect types that the inspection identified most often.

A derivation of concrete selection rules is skipped here. However, this can be done easily using the inspection defect profile; examples of concrete selection rules are shown in [2], [3], and some applied selection rules are shown in Section 4.5.

**Step 4 and 5: Selecting test cases and conducting the testing activities**

To evaluate the integrated inspection and testing approach and the stated assumptions, testing activities were performed without considering the inspection defect profile for the prioritization (however, the inspection defects were corrected before testing activities started). 40, respectively 42, similar test cases were applied during system test by the two testers, covering the main functionality of the tool, i.e., different kinds of reading support, the interaction of reading support and an artifact to be inspected, the generation of a report of the findings, and creating a checklist was tested. In addition, some explorative testing was performed by the tester that did not develop the tool.

Table 4. Test results from system testing

| Tested functionality | Number of test cases | | Number of defects found | | Defect ids | | Effort (min) | |
|---|---|---|---|---|---|---|---|---|
| | tester 1 | tester 2 | tester 1 | tester 2 | tester 1 | tester 2 | tester 1 | tester 2 |
| reading support: GIT | 3 | 3 | 1 | 1 | id1, id8* | id1 | 10 | 6 |
| reading support: SGIT | 3 | 3 | 0 | 1 | id9* | id1 | 7 | 6 |
| reading support: GC | 3 | 3 | 0 | 0 | id10* | | 7 | 6 |
| reading support: VID | 0 | 11 | 0 | 1 | id11* | id1 | 0 | 30 |
| reading support: CL | 1 | 1 | 0 | 0 | | | 3 | 2 |
| interaction | 15 | 8 | 5 | 2 | id2, id3, id4, id6, id7, id12* | id2, id3 | 33 | 21 |
| report generation | 1 | 1 | 1 | 0 | id5, id13* | | 15 | 10 |
| checklist creation | 16 | 10 | 1 | 0 | id4 | | 40 | 10 |

During the system test, seven additional defects regarding functionality were found by the two testers. Running the defined test cases took about 90, respectively 120 minutes. In addition, effort for explorative testing, test documentation, debugging, and correction was consumed, resulting in an overall test effort for both testers of about 14 hours. The distribution of defects with respect to functionality can be found in Table 4 (id1 – id7). Tester 1 found one defect (defect id 1) when testing the GIT reading support (which was inspected on the code level). However, this defect is independent of the concrete reading support. Tester 2 also found this defect when testing GITs, but also when testing the other tree-based reading model SGITs or VIDs. Furthermore, most of the defects occurred when testing the interaction between reading support and the artifact view. Two more defects were found when testing checklist creation and report generation. In addition, tester 1 found six more usability problems that were equally distributed (id8* – id13*), i.e., for almost each functionality tested, one usability problem was found.

### 3.6 Results of the Case Study and Lessons Learned

**RQ1.1:** Our first objective was to check whether the inspection defect information could be used to predict defect-proneness within code classes in order to focus unit testing activities. Unfortunately, the unit test activity could not be completed due to bad testability of the code and no new defects were found. Therefore, research question 1.1 could not be answered with respect to the unit level. Instead, the system test activity was used to analyze whether the code inspection results can provide valuable predictions for focusing system testing. Assumptions one and two were applied accordingly. We were aware that this prioritization would mean a different level of granularity, because for system tests it is not possible to address certain code classes; rather, they are used to address functionalities.

Five different kinds of reading support and three additional tool functionalities were tested and revealed that most of the defects were found in parts that had not been inspected. One functional defect was found when applying the GIT reading support (which was also inspected); however, this defect occurred independently of the concrete reading support and was also found when testing other kinds of reading support. Therefore, assumption one indicates a trend towards an appropriate prioritization, respectively prediction, of defect-proneness and might help in guiding system testing activities with reduced effort. Considering only the test execution effort, an effort reduction of between 8% and 23% could be achieved, depending on the concrete selection rules used. When defining a selection rule omitting the GIT test cases, this leads to the lowest reduction in the number of test cases, while all functional defects are found. Omitting SGITs as well, which are a very similar form of reading support, increases the saved effort. In addition, omitting test cases for checklists (i.e., low-complexity reading support), an effort reduction of up to 23% is achievable, with all functional defects still being found. However, the absolute numbers for conducting the tests are rather low and test derivation, documentation, and further activities are not considered here. Consequently, the numbers have to be treated with caution.

With respect to the evaluation of the study design, it is essential that appropriate testability exists in order to focus the testing activities on the same system level.

**RQ1.2:** Our second objective was to analyze the relationship between defect types identified in the inspection and during testing. Considering assumption three, many of those inspection defects classified as 'other' were documentation problems (e.g., missing explanation of a method, unclear description). Such kinds of defects affect the maintainability of the product under test and are not detectable with testing, since they do not influence the observable functionality. Regarding testing, six additional usability problems were found by a tester (e.g., bad readability of parts of reading support). Such kinds of problems can be identified if a graphical user interface is used during testing, but are usually not found during the inspection.

**Table 5.** ODC-classified defects from inspection and system testing

| ODC defect types | Inspection | Testing |
|---|---|---|
| algorithm / method | 53 | 2 |
| checking | 36 | 4 |
| function / class / object | 32 | 0 |
| assignment / initialization | 13 | 0 |
| relationship | 1 | 1 |
| timing / serialization | 0 | 0 |
| interface / o-o messages | 0 | 0 |
| other | 54 | 6 |
| Total | 189 | 13 |

In terms of maintainability and usability, it is rather easy to dedicate them to inspection respectively testing activities in order to find such problems. However, with respect to the ODC classification used for the inspection defects, detecting a relationship to defects found during the system test is difficult due to the difference in granularity between code defect types and system defect types. A post-mortem analysis of the seven functional defects found during testing revealed that most of them were classified as checking or algorithm / method defects, which fits exactly with the two defect types identified most often during inspections (see Table 5). Nevertheless, it is still unclear whether it is possible to derive system tests in a systematic manner that can address such kinds of problems and how this could be done. It might be possible that a defect classification, such as the ODC, is not suitable for guiding system test activities. An explorative study for identifying an appropriate defect classification would be necessary in this case. Finally, due to an uncompleted unit testing activity, no new insights regarding relationships between inspection defect types and testing defect types could be obtained on the unit level.

**RQ1:** To conclude the main results for research question one, a trend was found that testing activities might be guided based on inspection results, i.e., defect-prone parts and defect types could be predicted based on inspection defect data, and testing activities could be focused on certain parts and defect types. However, the quality of such focusing depends on the assumptions made in the given context. In our context, parts that had not been inspected contained additional defects that were found during testing. However, this can only be stated for defects found during system testing because unit testing could not be fully completed (which shows the importance of good testability). Therefore, appropriate testability is an inevitable prerequisite to performing a suitable evaluation, respectively application, of the integrated approach. An effort reduction for test case execution of up to 23% could be achieved when focusing on parts of the system, with the same level of quality being achieved. With respect to defect types, especially maintainability defects were found during inspections, while usability problems were found during testing.

The initial results of the preliminary study presented in this article together with results from the initial case study indicated a first trend towards the applicability of the approach and the potential for exploiting further synergy effects when integrating inspection and testing processes. As inspection and testing processes are established quality assurance activities that are widely applied in industry, and effort reduction is both a current and a future challenge, using inspection results to focus testing activities is a promising approach to be considered in order to address this challenge.

### 3.7 Threats to Validity

Next, we discuss what we consider to be the most relevant threats to validity.

**Conclusion validity:** The number of testers and the number of found test defects was low. One reason might be the low experience regarding testing. Consequently, no statistically significant data could be obtained.

**Construct validity:** To demonstrate the integrated approach, different assumptions were derived in our context. Four assumptions were used and analyzed regarding their suitability. However, more assumptions are reasonable and may lead to better or worse predictions. Finally, the selection of ODC was reasonable when focusing on the unit test level, but may not be suitable when using it for the system test.

**Internal validity:** The subjects selected for the inspection and for the testing activity may have influenced the number of defects that were found. However, the effect was slightly reduced by using checklists that focus an inspector on certain aspects in the code and by using equivalence partitioning, respectively addressing the main functionality, for the testing activity. Finally, defects could be classified differently.

**External validity:** The DETECT tool inspected and tested in the preliminary study is one example to which the integrated inspection and testing approach was applied. Few test defects were found that could be used for the analysis of the assumptions. Larger software, as typically developed by software companies, is expected to result in more test defects to be found. Assumptions have to be evaluated anew in each new environment, meaning that the conclusions drawn with respect to the used assumptions cannot be generalized directly. Finally, the results can only be transferred to a context where a comparable number of defects are found.

## 5 Summary and Outlook

To address the challenge of guiding testing activities, an integrated inspection and testing approach was presented that is used to predict defect-prone parts of a system and defect types of relevance in order to focus testing activities based on inspection results. This requires explicitly defined assumptions describing the relationships between inspection defects and testing defects. A preliminary case study was presented that analyzed four different assumptions in a given context. First trends could be seen regarding a prediction of defect-proneness. In addition, inspection and testing activities should be focused on those defect types that are most convenient to find, e.g., addressing maintainability problems during inspections, and usability problems during testing. It is worth noting that assumptions are probably not generally acceptable and thus, have to be re-evaluated in each new context in order to obtain evidence on them.

From an industry point of view, the integrated inspection and testing approach can lead to several benefits. Improvements in effectiveness and efficiency are often goals with respect to quality assurance activities. The approach might be applied in order to reduce test effort or to find more defects by focusing the available test effort on parts that are expected to be most defect-prone based on the inspection results. Furthermore, detailed knowledge about inspection and test relationships can lead to an improved overall quality assurance strategy and support the balancing of inspection and testing activities.

With respect to future work, an improvement of the approach and additional evaluations are planned. Based on the preliminary study and the evaluation of the study design, we got new insights into what is necessary with respect to a sound evaluation of the approach (e.g., appropriate testability). We recently began an analysis of inspection and test defect data from an industrial context in order to prove several of our assumptions as well as the potential for effort savings. The knowledge from this study will be incorporated.

Regarding the improvement of the integrated inspection and testing approach, more guidance on how to derive assumptions in a systematic manner should be defined. Furthermore, results from different inspection activities or inspections of only parts of an artifact might be used for guiding testing activities.

**Acknowledgments.** This work has been funded by the Stiftung Rheinland-Pfalz für Innovation project "Qualitäts-KIT" (grant: 925). We would also like to thank Stephan Kremer for tool development, all participants of the study, and Sonnhild Namingha for proofreading.

## References

[1] Cohen, J.: Best kept Secrets of Peer Code Review: Code Reviews at Cisco Systems, 63--87 (2006)

[2] Elberzhager, F. Eschbach, R. Muench, J.: Using Inspection Results for Prioritizing Test Activities. In: 21st International Symposium on Software Reliability Engineering, Supplemental Proceedings. 263--272 (2010) http://inspection.iese.de

[3] Elberzhager, F. Muench, J., Rombach, D., Freimut, F.: Optimizing Cost and Quality by Integrating Inspection and Test Processes. In: International Conference on Software and Systems Process, 3--12 (2011) http://inspection.iese.de

[4] Elberzhager, F., Eschbach, R., Muench, J.: The Relevance of Assumptions and Context Factors for the Integration of Inspections and Testing. In: 37th Euromicro Software Engineering and Advanced Application, Software Product and Process Improvement, in press (2011)

[5] Elberzhager, F., Eschbach, R., Rosbach, A., Münch, J.: Inspection and Test Process Integration based on Explicit Test Prioritization Strategies, IESE Report (2011)

[6] Elberzhager, F., Eschbach, R., Kloos, J.: Indicator-based Inspections: A Risk-oriented Quality Assurance Approach for Dependable Systems. In: Software Engineering 2010, GI-Edition Lecture Notes in Informatics, vol. 159, 105--116 (2010)

[7] Harrold, M.J.: Testing: A Roadmap. In: International Conference on Software Engineering. The Future of Software Engineering, 61--72 (2000)

[8] Orthogonal Defect Classification, IBM, http://www.research.ibm.com/softeng/ODC/ODC.HTM